\documentclass[12pt]{article}
\usepackage{graphicx}
\usepackage{amssymb}
\input epsf

\def\beq{\begin{equation}}
\def\eeq{\end{equation}}

\title{\bf $p_t$- dependence of the flow coefficients for pp collisions in the color string scenario.
Monte-Carlo simulations.}
\author{M.A. Braun$^a$, C. Pajares$^b$\\
$^a$ Dep. of High Energy physics,
 Saint-Petersburg State University, Russia\\
$^b$ Dep. of Particles, University of Santiago de Compostela, Spain}

\begin{document}
\maketitle
\begin{abstract}
In the color string picture with fusion and percolation
the dependence of the flow coefficients $v_n$ on the transverse momentum is
studied for pp collisions the  LHC energy respectively.  Monte-Carlo simulations are
used to locate simple strings and their fused clusters.
The results favorably agree with the CMS data  in the region $0.2 \le p_t\le 3.$ GeV/c
appropriate for the string scenario.
\end{abstract}

\section{Introduction}
One of the most impressing discoveries at LHC is observation of
strong azimuthal correlations in nucleus-nucleus collisions
~\cite{ref1,ref2,ref3}.
 It can be characterized by the non-zero
flow coefficients $v_n$ governing the correlation function of the azimuthal distribution of
secondaries as
\beq
C(\phi)=A(1+\Big(1+2\sum_{n=1}v_n\cos(n\phi)\Big).
\label{eq0}
\eeq

Several approaches tried to understand this effect.
The simplest approach was to match the initial anisotropic distribution of
participant nucleons with the following hydrodynamical evolution ~\cite{ref4}-\cite{ref6}. More sophisticated
treatments tried to relate the initial anisotropy with the preasymptotic partonic distributions in models
based on the Regge exchanges ~\cite{ref7}, in the Color Glass Condensate
framework ~\cite{ref8}-\cite{ref12} or in the color string scenario `\cite{ref13}-\cite{ref15}.

In the latter case both the flow coefficient and angle-rapidity correlations ("ridge")
were studied averaged over transverse momenta in the region 0--4 GeV/c.
In this paper we study the dependence of the flow coefficients on the transverse momenta,
as experimentally observed  in ~\cite{ref3}.
Color strings picture has been able to successfully explain many observable
phenomena in the soft dynamics domain. One expects it to be also applicable
to the flow problem in so far as one is dealing with relatively modest
transverse momenta.

Note that on very general grounds  in the color string framework the $p_t$ dependence of flow coefficients
in AA collisions exhibits a striking scaling behavior,
in a very good agreement with the data ~\cite{carlota1, carlota2}.

As experimental data show, the behavior of azimuthal correlations in pp collisions is very similar
to AA collisions provided one separates events of unusually high multiplicities. In the string picture this corresponds to
events in which the number of strings generated in a collision is much greater than the average.
Since in this scenario the dynamics totally depends on the number of strings and clusters formed in their fusion
then one indeed expects to eventually find a picture quite similar to AA collisions.
In our paper ~\cite{ref15}, in the framework of Monte-Carlo simulations, we indeed found that at the LHC energy with
50 formed strings as compared to the average 18 strings the azimuthal correlations are very similar to those for AA
collisions and to a good approximation agree with the experimental data in ~\cite{CMS1}.

In this paper we continue to study this phenomenon in a more detailed aspect. Namely we study the $p_t$ dependence
of the flow coefficients $v_n$ as calculated by Monte-Carlo simulations in the color string framework.
Our results can be compared with the  recent data on $v_2(p_t)$ published in \cite{CMS2}.
We find a rather good agreement,
which we interpret as confirmation of the validity of the color string approach to soft phenomena.

\section{String picture}

The color string model was proposed some time ago to describe
multiparticle production in the soft region. Its basic ideas can
be found in original papers and in a review~\cite{capella,kaidalov}.
Later the version of the model with string fusion and percolation~\cite{brapaj2},
in line with the ideas proposed in~\cite{biro}, was suggested.
Its application to the flow problem was developed in our previous paper
~\cite{ref14}. Here we only reproduce the main
points necessary to understand the technique.
It is assumed that in a high-energy collision between the partons of the
participants color strings are stretched, which may be visualized as
a sequence of $q\bar{q}$ pairs created from the vacuum or alternatively as
a strong gluonic field generated by the participant partons.
The strings are assumed to possess a certain finite dimension in
the transverse space related to confinement.
Each string then
breaks down in parts several times until its energy becomes of the order
of  GeV and it becomes an observed hadron. The number of strings
in the interaction area depends on the total available energy and
partonic structure of the colliding particles:
it grows with energy and atomic number. When the number of strings is
small they occupy a small part of the whole interaction area like drops
of liquid at considerable distance from one another. However when the number
of string grows they begin to overlap and fuse giving rise to strings with
more color and covering more space in the interaction area. At a certain
critical density strings begin to fuse forming clusters
 of the dimension comparable to that of the
interaction area (string percolation).
The basic assumptions which lie at the basis of the color string picture
are supported by its very successful application to
multiparticle production in the soft region. It describes
well  the multiplicity and transverse momentum distributions
and many other details of the particle spectra.
The color string picture has a certain similarity
with the saturation
(Color Glass Condensate or Glasma) models, where the dynamics is
explained by the classical gluon field stretched
between the colliding hadrons. The effective number of
independent color sources in string percolation can be put
in correspondence with the number of color flux tubes
in the Glasma.
It is found that they indeed have the same energy and number of participants
dependence. As a consequence predictions of both approaches for most
of the observables are similar.

It is assumed that strings
decay into particles ($q\bar{q}$ pairs) by the well-known mechanism for
pair creation in a strong
electromagnetic field. In its simplest version,
the particle distribution at the moment of its production by the string is
\beq
P(p,\phi)=Ce^{-\frac{p_0^2}{T}},
\label{prob}
\eeq
where $p_0$ is the particle initial transverse momentum,
$T$ is the string tension (up to an irrelevant numerical coefficient) and
$C$ is the normalization factor.
However, as proposed in~\cite{ref14}, $p_0$ is different from the
observed particle momentum $p$ because
the particle has to pass through the  fused string area and emit gluons
on its way out. So in fact in Eq. (\ref{prob}) one has to consider $p_0$
as a function of $p$ and path length $l$ inside the nuclear overlap:
$p_0=f(p,l(\phi))$ where $\phi$ is the azimuthal angle. Note that Eq. (\ref{prob})
describes the spectra only at very soft $p_0$. To extend its validity to
higher momenta one may use the idea that the string tension fluctuates, which
transforms the Gaussian distribution into the thermal one~\cite{bialas,deupaj}:
\beq
P(p,\phi)=Ce^{-\frac{p_0}{t}}
\label{probb}
\eeq
with temperature $t=\sqrt{T/2}$
To describe the energy loss of the parton due to gluon emission
one may use the corresponding QED picture for a charged particle  moving
in the external electromagnetic field \cite{nikishov}. This leads to
the quenching formula~\cite{ref14}
\beq
p_0(p,l)=p\Big(1+\kappa p^{-1/3}T^{2/3}l\Big)^3,
\label{quench1}
\eeq
with the quenching coefficient $\kappa$ to be taken from the experimental
data. We adjusted $\kappa$  to
give the experimental value for the coefficient $v_2$
in mid-central Pb-Pb collisions at 5-13 TeV GeV, integrated over the
transverse momenta, which gives $\kappa=0.48$.

Of course the possibility to use electrodynamic formulas for
the chromodynamic case may raise certain doubts. However in
~\cite{mikhailov} it was found that at least in the $N=4$ SUSY
Yang-Mills case the loss of energy of a colored charge moving
in the external chromodynamic field was given by essentially the same
expression as in the QED.

\section{Calculations}
The general scheme of calculations repeats the one presented in
our previous papers dedicated to flow coefficients~\cite{ref14,ref15}.
So here we only briefly describe the main points.

For a particular event, that is, for a given string configuration with a fixed azimuthal angle $\phi_0$ of the
impact parameter vector,
the inclusive cross-section to produce a particle with a given transverse momentum and azimuthal angle $p_t$ and $\phi$ at fixed rapidity is
\[
I^e(p_t,\phi)=A^e(p_t)+2\sum_{n=1}\Big(B_n^e(p_t)\cos n(\phi-\phi_0)+
C^e_n(p_t)\sin n(\phi-\phi_0)\Big)\]\beq
=
A^e(p_t)\Big[1+2\sum_{n=1}\Big(a_n^e(p_t)\cos n\phi+
b_n^e(p_t)\sin n\phi\Big)\Big].
\label{eq1}
\eeq
The flow coefficients for this event are given by
\beq
v_n^e(p_t)=\Big[(a_n^e)^2(p_t)+(b_n^e)^2(p_t)\Big]^{1/2}.
\label{vn}
\eeq
The experimentally observed flow coefficients, $v_n(p_t)$, are obtained after
averaging over different events
\beq
v_n(p_t)= \langle v_n^e(p_t) \rangle.
\label{vnexp}
\eeq

Integrating $I^e(p_t,\phi)$ over $p_t$ in a certain interval one obtains the integrated
inclusive distribution for the event depending only on $\phi$.
It can also be presented as (\ref{eq1})
\beq
I^e(\phi)=
A^e\Big[1+2\sum_{n=1}\Big(a_n^e\cos n\phi+
b_n^e\sin n\phi\Big)\Big]
\label{eq11}
\eeq
and the integrated flow coefficients $v_n$ are obtained after averaging
\beq
v_n=\Big<\Big[(a_n^e)^2+(b_n^e)^2\Big]^{1/2}\Big>.
\eeq
The observed integrated single inclusive cross section, obtained after averaging (\ref{eq11})
does not depend on $\phi$:
\beq
I=\langle I^e(\phi)\rangle=\langle A^e\rangle.
\eeq

If we neglect correlations for emissions from a single string, then the double
inclusive cross-section for an event, integrated over the transverse momenta of both observed particles,
 is just the product of two $I^e(\phi)$, Eq. (\ref{eq11}).
Its averaging over events gives the experimentally observed double inclusive
cross-section
\beq
I_2(\phi_{12})= \Big\langle {A^e}^2\Big(1+2\sum_n  {v_n^e}^2 \cos n\phi_{12}\Big\rangle,
\label{iexp}
\eeq
where $\phi_{12}=\phi_1-\phi_2$.
The correlation function
 \[
C(\phi_{12})=\frac{I_2(\phi_{12})}{I^2}-1  \ .
\]
is then
\beq
C(\phi_{12})=\frac{1}{I^2}\Big\langle 2A^2\sum_{n=1}{v_n^e}^2\cos n\phi_{12}+A^2-<A>^2\Big\rangle
\eeq

In the Monte-Carlo technique, in the framework of the color string approach, at each simulation one throws strings onto
the interaction area and determines the event distribution $I^e(p_t\phi)$ and the related event flow coefficients $v_n^e(p_t)$.
Successive simulations and the following averaging give the observed flow coefficient $v_n(p_t)$, the integrated ones $v_n$
and correlation
function $C(\phi_{12})$.
Note that the experimental data
mostly assume multiplying $C(\phi)$ by the multiplicity minus unity,
that is normalizing the correlation function according to
\beq
 C(\phi) \to  \langle A-1 \rangle  C(\phi).
\eeq
Presenting our results we use this normalization.

Following these lines
a Monte-Carlo code
was developed for proton-proton collisions.
Each colliding nucleon was presented as a disk
of the typical nucleon radius 0.8 fm and with the matter distributed inside according to
the Gaussian density.

\begin{figure}
\includegraphics[width=8.2 cm]{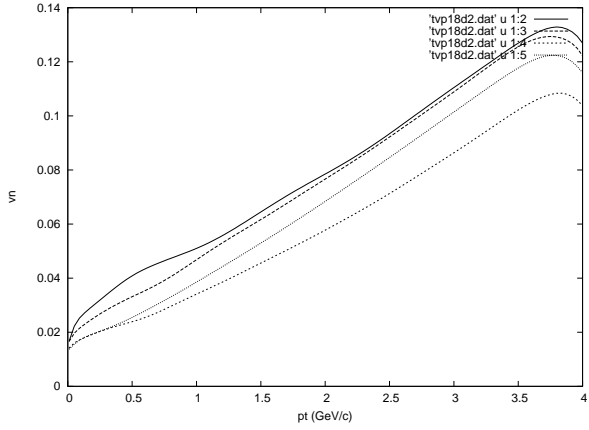}
\caption{Flow coefficients $v_n(p_t)$ for pp collisions at 5-13 TeV
with average multiplicity}
\label{fig7}
\end{figure}

\begin{figure}
\includegraphics[width=8.2 cm]{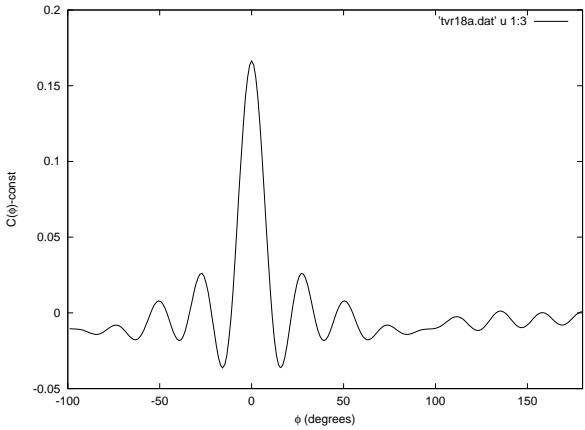}
\caption{Correlation coefficient $C(\phi)$ for pp
collisions at 5-13 TeV with average multiplicity}
\label{fig8}
\end{figure}

 From  ~\cite{BDMP} at
 5-13  TeV the average number of strings is approximately 18.
Our results for  energies in this interval are very similar.
Calculations have shown that for such small
number of strings fluctuations are quite strong, so that reliable
results can be obtained after no less that 1000 simulations.
In Fig. \ref{fig7} we show
coefficients $v_n(p_t)$ as functions of $p_t$.
Approximating the behavior of $v_n(p_t)$ with the growth of $p_t$ as $A\, p_t^{\alpha_n}$
we find $\alpha_n=0.59, 0.70, 0.77$ and 0.83 for $n=2,3,4,5$ respectively.
The flow coefficients $v_n$ corresponding to integration in the interval $p_t<4$ GeV/c
are shown in Table 1. in columns 2-4 . In the second column we present $v_n$ for minimum bias events with the temperature distribution (\ref{probb}).
It is remarkable that they do not diminish with $n$
in contrast to the events with triple multiplicity (third and fourth columns).
In Fig. \ref{fig8} we present the correlation coefficient $C(\phi)$
for these minimal-bias events.
All the $\phi$ dependence is found to be collimated to quite small angles
$\phi\leq 10^0$. Note that
in this case the constant term $(<A^2>-<A>^2)/<A>^2$ dropped in Fig. \ref{fig8}
is of the order unity, so that the ridge turns out to be only a small
ripple against a constant background.

\begin{table}
\caption{Integrated flow coefficients $v_n$ for minimum bias events
and events with a triple multiplicity and with temperature and Schwinger
distributions in $p_t$}
\begin{center}
  \begin{tabular}{|r|r|r|r|}

  \hline
    n&   min.bias&    temperature&   Schwinger\\\hline
   1&   0.2528E-01&   0.1901E-01&    0.1878E-01\\
   2&   0.2884E-01&   0.2308E-01&    0.2240E-01\\
   3&   0.2470E-01&   0.1668E-01&    0.1648E-01\\
   4&   0.1941E-01&   0.9273E-02&    0.9024E-02\\
   5&   0.2023E-01&   0.1130E-01&    0.1115E-01\\
   6&   0.1869E-01&   0.2775E-02&    0.2671E-02\\
   7&   0.2007E-01&   0.5928E-02&    0.5855E-02\\
   8&   0.1854E-01&   0.1343E-02&    0.1254E-02\\
   9&   0.1801E-01&   0.3718E-02&    0.3669E-02\\
  10&   0.1824E-01&   0.7131E-03&    0.6819E-03\\
  11&   0.1912E-01&   0.4231E-02&    0.4177E-02\\
  12&   0.1828E-01&   0.1558E-02&    0.1477E-02\\
  13&   0.2108E-01&   0.3833E-02&    0.3787E-02\\
  14&   0.2036E-01&   0.3478E-03&    0.3344E-03\\
  15&   0.1633E-01&   0.2464E-02&    0.2437E-02\\
  16&   0.1511E-01&   0.4720E-03&    0.4425E-03\\\hline
 \end{tabular}

\end{center}
\end{table}

\begin{figure}
\includegraphics[width=8.2 cm]{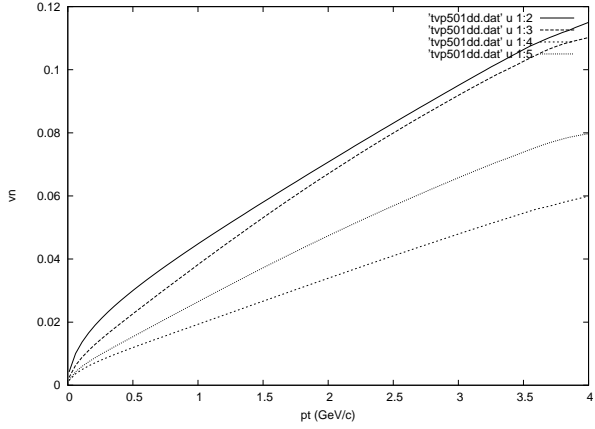}
\caption{Flow coefficients $v_n(p_t)$ for pp collisions at 5-13 TeV
with triple multiplicity}
\label{fig9}
\end{figure}

\begin{figure}
\includegraphics[width=8.2 cm]{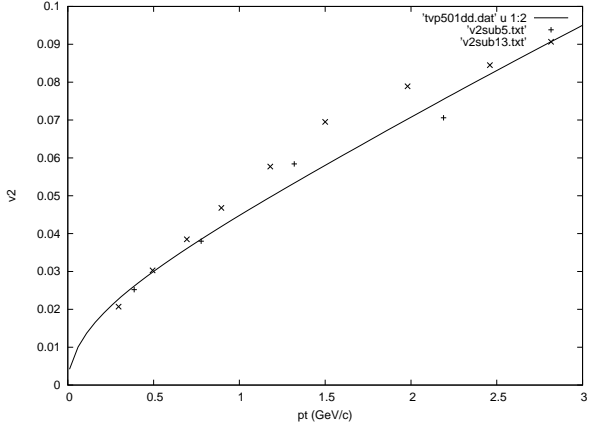}
\caption{Flow coefficient $v_2(p_t)$ for pp
collisions with triple multiplicity
compared to  the experimental data at 5 TeV (lower)and 13 TeV (higher) from ~\cite{CMS2}}
\label{fig10}
\end{figure}

\begin{figure}
\hspace*{2.5 cm}
\includegraphics[width=8.2 cm]{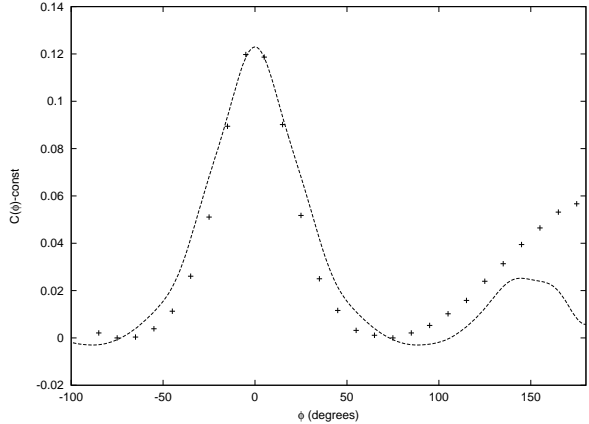}
\caption{Correlation coefficient $C(\phi)$ for pp
collisions at 5 and 13 TeV with triple multiplicity
compared to the experimental data from ~\cite{CMS1}
(with the ZYAM procedure at positive $\phi$)}
\label{fig11}
\end{figure}

Next, following the experimental observations, we studied  rare cases in which
the multiplicity is three or more times greater than the average.
The maximal number of string is then found to be 50. The resulting
dependence of the flow coefficients $v_n(p_t)$  on $p_t$ is
shown in Fig. \ref{fig9}.
Taking again $v_n(p_t)\simeq A\,p_t^{\alpha_n}$ we now have
$v_n=0.65,0.79.0.80$ and 0.82 for $n=2,3,4,5$.
So
the rise of $v_n$ with $p_t$ on the whole turns out to be similar to the minimum bias events.
In Fig. \ref{fig10}
we compare our results for $v_2(p_t)$ with the experimental data from
 ~\cite{CMS2}.
measured for two energies 5 and 13 GeV/c (which are not very different). As we see our results
agree rather well with both sets of data.
The flow coefficients $v_n$ corresponding to integration over $p_t>4$ GeV/c are
shown in Table 1. in the third column.
The correlation coefficient is presented in  Fig. \ref{fig11}
 As one observes
the correlation coefficient then becomes quite similar to the one
in AA collisions in good agreement with the experimental findings in ~\cite{CMS1}. Still
the dropped constant term is again of the order unity, so that the
ridge stands on a large constant pedestal.

\begin{figure}
\includegraphics[width=8.2 cm]{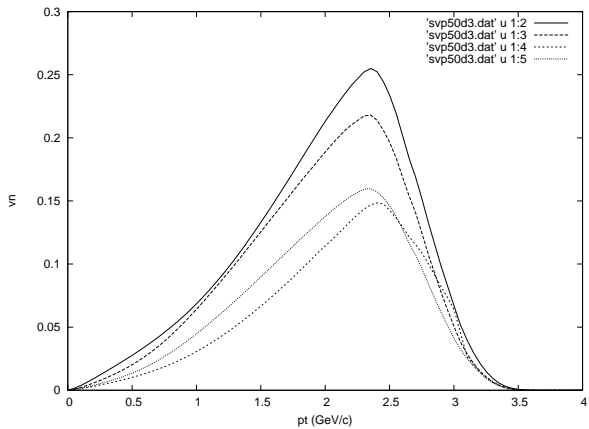}
\caption{Flow coefficients $v_n$ for pp collisions at 5-13 TeV
with triple multiplicity, calculated with the Schwinger distribution (\ref{prob}).}
\label{fig12}
\end{figure}

Just for comparison we present the coefficients $v_n(p_t)$ calculated with the
original Schwinger distribution (\ref{prob}) in Fig \ref{fig12}. The found
coefficients  rise to their  maximum  much faster
than observed experimentally.
The corresponding integrated $v_n$ are shown in Table in the fourth column.
Curiously they  are practically the same as for the temperature distribution, which obviously testifies that
the Schwinger distribution describes emission at  low momenta dominating the integrated quantities quite well.
It fails to work at medium momenta, which is evident from the comparison of Figs. \ref{fig9} and \ref{fig12}.
As a result the correlation coefficient calculated with the Schwinger distribution is practically
identical with the one calculated with the temperature distribution shown in Fig. \ref{fig11}.

\section{Conclusions}
Within the color string scenario via Monte-Carlo simulation the flow coefficients for proton-proton collisions
at 5-13 TeV were calculated both as functions of the transverse momentum and integrated over these in the interval
$0<p_t<4$ GeV/c. The found $v_2(p_t)$ and correlation coefficient favorably agree with the recent experimental data.
It remains to be seen in which degree the obtained agreement depends on the assumed form of quenching.
This may shed light on the behavior of $v_n(p_t)$ in AA collisions, which experimentally shows a remarkable
simple dependence of $\alpha_n=n/3$ in the parametrization $v_n(p_t)\propto p^{\alpha_n}$ ~\cite{ALICE}.

\section{Acknowledgements}
M.A.B. appreciates hospitality and financial support of the University of Santiago de Compostela, Spain.
C.P. thanks the grant Maria de Maeztu Unit of Excelence of Spain and the support
Xunta de Galicia. This work was partially done under the project EPA 2017-83814-P
of Ministerio Ciencia, Tecnologia y Universidades of Spain.

%



\begin{thebibliography}{100}
\bibitem{ref1} S.Afanasiev {\it et al}, PHENIX collab., Phys.Rev. {\bf C 80} (2009) 024909 [nucl-ex/ 0905.1070]
%
\bibitem{ref2} R.Aamodt {\it et al}, ALICE collab., Phys. Rev. Lett.  {\bf 107} (2011) 032301 [nucl.ex/1105.3865]
%
\bibitem{ref3} A.Adare {\it et al} PHENIX collab., Phys. Rev. Lett. {\bf 107} (2011) 252301 [nucl-ex/1105.39.28]
%
%
\bibitem{ref4} H.Holopainen, H.Niemi and K.J.Eskola, Phys.Rev. {\bf C 83}
(2011) 034901 [hep-ph/1007.0368].
%
\bibitem{ref5} H.Petersen,G-Y.Quin, S.A.Bass and B.Mueller,
Phys.Rev, {\bf C 82} (2010) 041901,064903 [nucl-th/1008.0625;1009.1847].
%
\bibitem{ref6} Zhi Qiu and U.Heinz, Phys.Rev.{\bf C 84} (2011) 024911
[nucl-th/1104.0650].
%
\bibitem{ref7} K.G.Boreskov, A.B.Kaidalov, O.V.Kancheli, Eur. Phys. J {\bf C 58} (2008) 445
%
\bibitem{ref8} A.Dumitru, K.Dusling, F.Gelis, J.Jalilian-Marian, T.Lappi and
R.Venugopalan, Phys. Lett. {\bf B 697} (2011) 21.
%
\bibitem{ref9} S.Gavin, G.Moschelli, Phys. Rev. {\bf C 85} (2012) 014905.
%
\bibitem{ref10} K.Dusling, T.Venugopalan, Phys. Rev. Lett. {\bf 108}
(2013) 262001,(arXiv:1201.2658 [hep-ph]).
%
\bibitem{ref11} K.Dusling, T.Venugopalan, Phys. Rev. {\bf D87}
(2013) 094014, (arXiv:1302.7018 [hep-ph]).
%
\bibitem{ref12} A.Bzdak, B.Schenke, P.Tribedy, R,Venugopalan, Phys. Rev.
{\bf C 87} (2013) 064906.
%
%
%
%
%
%
%
%
%
%
%
%

\bibitem{ref13} M.A.Braun, C.Pajares, Eur. Phys. J. {\bf C 71} (2011) 1558
%
\bibitem{ref14} M.A.Braun, C.Pajares, V.V.Vechernin, Nucl. Phys. {\bf A 906} (2013) 14
%
\bibitem{ref15} M.A.Braun, C.Pajares, V.V.Vechernin, Eur. Phys. J. {\bf A 51} (2015) 44
%
\bibitem{carlota1} C.Andres, J.Dias de Deus, A.Moscoso, C.Pajares, C.Salgado, Phys. Rev {\bf C 92} (2015) 034961
%
\bibitem{carlota2} C.Andres, M.A.Braun, C.Pajares, Eur. Phys. J. {\bf A 53} (2017) 41
%
\bibitem{CMS1} CMS collab. JHEP 1009 (2010) 091; arXiv:1009.4122 [hep-ex]
%
\bibitem{CMS2} CMS collab. Phys. Lett. {\bf 765} (2017) 193; arXiv:1606.06198 [nucl-ex]
%
%
%
%
%
%
%
%
%
%
%
%
%
\bibitem{capella} A.Capella, U.P.Sukhatme, C.-I.Tan and J.Tran Thanh Van,
Phys. Lett. {\bf B 81} (1979), 68; Phys. Rep. {bf 236} (1994) 225.
%
\bibitem{kaidalov} A.B.Kaidalov and K.A.Ter-Martirosyan, Phys. Lett.
{\bf B 117} (1982) 247.
%
\bibitem{brapaj2} M.A.Braun and C.Pajares, Phys. Lett. {\bf B 287} (1992) 154;
Nucl. Phys. {\bf B 390} (1993) 542; ibid 559.
%
\bibitem{biro} T.S. Biro, H.B. Nielsen and J.Knoll,
Nucl. Phys. {\bf B 245} (1984) 449.
%
%
\bibitem{bialas} A.Bialas, Phys. Lett. {\bf B 466} (1999) 301.
%
\bibitem{deupaj} J.Dias de Deus and C.Pajares, Phys. Lett.{\bf B 642} (2006) 455.
%
\bibitem{nikishov} A.I.Nikishov, Nucl. Phys. {\bf B 21} (1970) 346.
%
\bibitem{mikhailov} A.Mikhailov, arXiv:hep-th/0305196.
%
%
\bibitem{tarn} T.J. Tarnowsky, B.K. Srivastava and R.P. Scharenberg (for the STAR {\it collab}),
Nucleonica {\bf 51S3} (2006) S109, (arXiv:nucl-ex/0606019).
%
\bibitem{DDD} J.Dias de Deus, A.S.Hirsch, C.Pajares, R.P.Scharenberg
and B.K.Srivastava, Eur. Pgys. J {\bf C 72} (2012) 2123,
(arxiv:1106.4271 [nucl-ex]).
%
\bibitem{ALICE} ALICE collab., arXiv:1804.02944 [nucl-ex]
%
\bibitem{BDMP} I.Bautista, J. Dias de Deus, J.G.Milhano, C.Pajares,
Phys. Lett, {\bf B 715} (2012) 230, (arxiv:1204.1457 [nucl-th]).
%
\end{thebibliography}
\end{document}